\documentclass[times,10pt,twocolumn]{article} 
\usepackage{latex8}
\usepackage{times}

\begin{document}
\title{Improvements to the Psi-SSA Representation}
\author{Fran\c{c}ois de Ferri\`ere \\
       ST Microelectronics\\
       12 rue Jules Horowitz, 38000 Grenoble, France\\
       francois.de-ferriere@st.com\\
}
\maketitle
\thispagestyle{empty}



%

\begin{abstract}
Modern compiler implementations use the Static Single Assignment
representation~\cite{CFR+91} as a way to efficiently implement
optimizing algorithms. However this representation is not well adapted
to architectures with a predicated instruction set. The $\psi$-SSA
representation was first introduced in~\cite{Stou01} as an extension
to the Static Single Assignment representation. The $\psi$-SSA
representation extends the SSA representation such that standard SSA
algorithms can be easily adapted to an architecture with a fully
predicated instruction set. A new pseudo operation, the $\psi$
operation, is introduced to merge several conditional definitions into
a unique definition.

This paper presents an adaptation of the $\psi$-SSA representation to
support architectures with a partially predicated instruction set. The
definition of the $\psi$ operation is extended to support the
generation and the optimization of partially predicated code. In
particular, a predicate promotion transformation is introduced to
reduce the number of predicated operations, as well as the number of
operations used to compute guard registers. An out of $\psi$-SSA
algorithm is also described, which fixes and improves the algorithm
described in~\cite{Stou01}. This algorithm is derived from the out of
SSA algorithm from Sreedhar et al.~\cite{VC+99}, where the definitions
of liveness and interferences have been extended for the $\psi$
operations. This algorithm inserts predicated copy operations to
restore the correct semantics in the program in a non-SSA form.

The $\psi$-SSA representation is used in our production compilers,
based on the Open64 technology, for the ST200 family processors. In this
compiler, predicated code is generated by an if-conversion algorithm
performed under the $\psi$-SSA representation~\cite{Stou04, Bru06}.
\end{abstract}

\Section{Introduction}

The Static Single Assignment representation was introduced
in~\cite{CFR+91} and is now widely used in modern compilers. The SSA
representation has proven to be a very efficient internal compiler
representation for performing various optimizations on scalar
variables. In this representation, each definition of a scalar
variable is renamed into a unique name, and variable uses are renamed
to refer to these new definition names or to special $\phi$
instructions that are introduced to merge values coming from different
control-flow paths. Most of the standard optimization algorithms have
been adapted to this representation, such as constant
propagation~\cite{WZ91}, dead-code elimination~\cite{morgan98},
induction variables optimization~\cite{Wolfe92}, and partial redundancy
elimination~\cite{CCK+97}. These algorithms usually perform equally well or
even better than their original versions on a non-SSA
representation. However, these algorithms are more difficult to adapt
in presence of aliased variables, partial definitions or
conditional definitions. To overcome these difficulties, some
extensions to the SSA representation have already been proposed, such
as the HSSA representation~\cite{ICCC::ChowCLLS1996} for aliases with
pointers, the Array SSA form~\cite{KS98} for array variables and the
$\psi$-SSA representation~\cite{Stou01} to handle conditional
definitions.

In this document we present an extension of the $\psi$-SSA
representation for partially predicated architectures. The first
section will present theoretical and practical aspects of the
$\psi$-SSA representation. The second section will then describe the
adaptation of the $\psi$-SSA representation to the context of partial
predication. The third section will present an out of SSA algorithm
for the $\psi$-SSA representation, for both full and partial
predication. This algorithm improves the algorithm described in the
original $\psi$-SSA paper and also fixes some errors. In the fourth
section we will present some results we have on our production
compiler for one of the ST200 family processors.

\Section{The Psi-SSA representation}

The $\psi$-SSA representation was developed to extend the SSA
representation with support for predicated operations. In the SSA
representation, each definition of a variable is given a unique name,
and new pseudo definitions are introduced on $\phi$ instructions to
merge values coming from different control-flow paths. In this
representation, each definition is an unconditional definition, and
the value of a variable is the value of the expression on the unique
assignment to this variable. This essential property of the SSA
representation does not any longer hold when definitions may be
conditionally executed. When the definition for a variable is a
predicated operation, this operation is executed depending on the
value of a guard register. As a result, the value of the variable
after the predicated operation is either the value of the expression
on the assignment if the predicate is true, or the value the variable
had before this operation if the predicate is false. We need a way to
express these conditional definitions whilst keeping the static single
assignment property.

Predicated operations can be used to replace code that contains
control-flow edges by straight line code containing predicated
operations. Such a transformation is performed by an if-conversion
optimization~\cite{Fang96, Bru06}. A simple example of if-conversion
is given in figure~\ref{fig:op_pred}. In the rest of this paper, we
use the notation {\tt p?~<exp>} to say that {\tt <exp>} is executed
only if the predicate {\tt p} is TRUE.

\begin{figure}
\begin{center}
\footnotesize
\begin{tabular}{llll}
${\tt if (p)}$        & & &\\
${\tt\ \ \ \ \ a = op1;}$ & \ \ \ \ \  & ${\tt p?}$ & ${\tt a = op1;}$ \\
${\tt else}$          & & & \\
${\tt\ \ \ \ \     b = op2;}$ & \ \ \ \ \  & ${\tt \overline{p}?}$ & ${\tt b = op2;}$ \\
${\tt x = Phi(a, b)}$ & & & ${\tt x = Psi(a, b)}$ \\
\end{tabular}
\caption{$\psi$-SSA representation}
\label{fig:op_pred}
\end{center}
\end{figure}

In the $\psi$-SSA representation, $\psi$ operations are added to the
SSA representation. $\psi$ operations are for predicated definitions
what $\phi$ operations are for definitions on different control-flow
edges. A $\psi$ operation merges values that are defined under different
predicates, and defines a single variable to represent these different
values.

In the SSA representation, $\phi$ operations are placed at
control-flow merge points where each argument flows from a different
incoming edge. In the $\psi$-SSA representation, on a $\psi$
operation, all the incoming edges of a $\phi$ operation
are merged into a single execution path, and each argument is now
defined on a different predicate.

In figure~\ref{fig:op_pred}, variables {\tt a} and {\tt b} were
initially the same variable. On the left-hand side, the SSA construction
renamed the two definitions of this unique variable into two different
names, and introduced a new variable {\tt x} defined by a $\phi$
operation to merge the two values coming from the different
control-flow paths. On the right-hand side, an if-conversion algorithm
transformed this code to remove the control-flow edges. It introduced
predicated operations for the definitions of the variables {\tt a} and
{\tt b} and turned the $\phi$ operation into a $\psi$ operation. Each
argument of the $\psi$ operation is defined by a predicated
operation. The intersection of the domain of the two predicates is
empty and the value of the $\psi$ operation is given by one or the
other of its arguments, depending on the value of the predicate.

The $\psi$ operations can also represent cases where variables are
defined on predicates that are computed from independent
conditions. This is illustrated in figure~\ref{fig:non_disjoint_pred},
where the predicates {\tt p} and {\tt q} are independent. During the
SSA construction a unique variable was renamed into the variables {\tt
a}, {\tt b} and {\tt c} and the variables {\tt x} and {\tt y} were
introduced to merge values coming from different control-flow
paths. In the non-predicated code, there is a control-dependency
between {\tt x} and {\tt c}, which means the definition of {\tt c}
must be executed after the value for {\tt x} has been computed. In the
predicated form of this example, there are no longer any control
dependencies between the definitions of {\tt a}, {\tt b} and {\tt
c}. A compiler transformation can now freely move these definitions
independently of each other, which may allow more optimizations to be
performed on this code. However, the semantics of the original code
requires that the definition of {\tt c} occurs after the definitions
of {\tt a} and {\tt b}. We use the order of the arguments in a $\psi$
operation to keep the information on the original order of the
definitions.
We take the convention that the order of the arguments in a $\psi$
operation is, from left to right, equal to the order of their
definitions, from top to bottom, in the control-flow dominance tree of
the program in a non-SSA representation. This information is needed to
maintain the correct semantics of the code during transformations of
the $\psi$-SSA representation and when reverting the code back to a
non $\psi$-SSA representation.

\begin{figure}
\begin{center}
\footnotesize
\begin{tabular}{llll}
${\tt if (p)}$        & & &\\
${\tt\ \ \ \ \ a = 1;}$ & \ \ \ \ \  & ${\tt p?}$ & ${\tt a = 1;}$ \\
${\tt else}$          & & & \\
${\tt\ \ \ \ \     b = -1;}$ & \ \ \ \ \  & ${\tt \overline{p}?}$ & ${\tt b = -1;}$ \\
${\tt x = Phi(a, b)}$ & & & ${\tt x = Psi(a, b)}$ \\
${\tt if (q)}$        & & &\\
${\tt\ \ \ \ \ c = 0;}$ & \ \ \ \ \  & ${\tt q?}$ & ${\tt c = 0;}$ \\
${\tt y = Phi(x, c)}$ & & & ${\tt y = Psi(a, b, c)}$ \\
\end{tabular}
\caption{$\psi$-SSA with non-disjoint predicates}
\label{fig:non_disjoint_pred}
\end{center}
\end{figure}

With this definition of the $\psi$-SSA representation, conditional
definitions on predicated code are now replaced by unconditional
definitions on $\psi$ operations. Usual algorithms that perform
optimizations or transformations on the SSA representation can now be
easily adapted to the $\psi$-SSA representation, without compromising
the efficiency of the transformations performed. Actually, within the
$\psi$-SSA representation, predicated definitions behave exactly the
same as non predicated ones for optimizations on the SSA
representation. Only the $\psi$ operations have to be treated in a
specific way. As an example, the constant propagation algorithm
described in~\cite{WZ91} can be easily adapted to the $\psi$-SSA
representation. In this algorithm, the only modification is that
$\psi$ operations have to be handled with the same rules as the $\phi$
operations. We have also ported dead code elimination~\cite{morgan98}
and global value numbering~\cite{Cli95} algorithms to this
representation, and we expect that partial redundancy
elimination~\cite{CCK+97}, and induction variable
analysis~\cite{Wolfe92} should be easy to adapt.

In addition to standard algorithms that can now be easily adapted to
$\psi$ operations and predicated code, a number of additional
transformations can be performed on the $\psi$ operations. These
transformation are $\psi$-inlining, $\psi$-reduction and
$\psi$-projection, they are described in detail
in~\cite{Stou01}. $\psi$-inlining will recursively replace in a $\psi$
operation an argument that is defined by another $\psi$ operation by the
arguments of this second $\psi$ operation. $\psi$-reduction will remove from
a $\psi$ operation an argument whose value will always be overridden by
arguments on its right in the argument list, because the domain of the
predicate associated with this argument is included in the union of
the domains of the predicates associated with the arguments on its
right. $\psi$-projection will create from a $\psi$ operation new
$\psi$ operations for uses in operations guarded by different
predicates. Each new $\psi$ operation is created as the projection on
a given predicate of the original $\psi$ operation . In this new
$\psi$ operation, arguments whose associated predicate has a domain
that is disjoint with the domain of the predicate on which the
projection is performed actually contribute no value to the $\psi$
operation and are then removed.


\Section{Psi-SSA and partial predication}

In the original paper on $\psi$-SSA we only considered the use of
$\psi$-SSA for a fully predicated processor. We describe here how this
representation has been modified to be used for a processor with a
partially predicated instruction set.

In a partially predicated instruction set, only a subset of the
instruction set of the targeted processor supports a predicate
operand. For example, the instruction set may support only a
conditional {\tt move} instruction. It can also include more specific
instructions such as a {\tt select} instruction. A {\tt select}
instruction takes two arguments and a guard register, and assigns the
value of one or the other of its arguments into a variable, depending
on the value of the guard register.

The only impact of partial predication on the $\psi$-SSA
representation is that when a $\psi$ operation is created as a
replacement for a $\phi$ operation, during if-conversion for example,
some of its arguments may be defined by operations that cannot be
predicated. A preliminary condition is that the $\psi$ operation can
be created only if these non-predicated arguments can be safely
speculated, which means executed under some conditions where they
would not have been executed otherwise. Although these definitions are
speculated, their values were only meaningful under a given predicate
in the original code. The information on this predicate must be kept
in some way.

\begin{figure*}
\begin{center}
\footnotesize
\begin{tabular}{llllll}
${\tt if (p)}$                & & & & &\\
${\tt\ \ \ \ \ a = ADD\ i,1;}$ & \ \ \ \ \ \ \ \  & & ${\tt a = ADD\ i,1;}$   & \ \ \ \ \ \ \ \  & ${\tt a = ADD\ i,1;}$ \\
${\tt else}$                  & \ \ \ \ \ \ \ \  & ${\tt p?}$ & ${\tt c = a}$ & \ \ \ \ \ \ \ \  & \\
${\tt\ \ \ \ \ b = ADD\ i,2;}$ & \ \ \ \ \ \ \ \  & & ${\tt b = ADD\ i,2;}$   & \ \ \ \ \ \ \ \  & ${\tt b = ADD\ i,2;}$ \\
                            & \ \ \ \ \ \ \ \  & ${\tt \overline{p}?}$ & ${\tt d = b}$ & \\
${\tt x = Phi(a, b)}$         & \ \ \ \ \ \ \ \  & & ${\tt x = Psi(c, d)}$  & \ \ \ \ \ \ \ \  & ${\tt x = Psi(p?a, \overline{p}?b)}$\\
\\
\\
\multicolumn{2}{l}{${\tt{\bf a)\ before\ if-conversion}}$} & \multicolumn{2}{l}{${\tt{\bf b)\ conditional\ moves}}$} & \multicolumn{2}{l}{${\tt{\bf c)\ extended\ Psi\ operation}}$}\\
\end{tabular}
\caption{Psi-SSA for partial predication}
\label{fig:psi_partial}
\end{center}
\end{figure*}

Figure~\ref{fig:psi_partial} shows an example where some code with
control-flow edges was transformed into a linear sequence of
instructions. In this example, the {\tt ADD} operation cannot be
predicated.

In figure~\ref{fig:psi_partial} b), we have introduced predicated move
operations, so that $\psi$ operations still have the definitions of
their arguments being predicated, while allowing an if-conversion
transformation to be performed even on operations that cannot be
predicated. In the case where conditional move operations are not
available on the target processor, when leaving the $\psi$-SSA
representation, these operations, along with the $\psi$ operation,
will be replaced by other operations available on the target
processor, such as a {\tt select} instruction for example. The main
disadvantage of this solution is that the semantics of the initial
$\phi$ operation is now expressed by three operations. These
operations will have to be treated all together during transformations
on the $\psi$ operations, and in particular when reverting the code
back to non $\psi$-SSA representation.

In figure~\ref{fig:psi_partial} c), we chose to express these {\tt
conditional move} operations directly in the $\psi$ operation, by
means of a predicate associated with each argument of the $\psi$
operation. With this representation, the information represented in
the $\phi$ operation by the control-flow edges is now present in the
$\psi$ operation by means of predicates.

In the general case, the definition of a variable can be
predicated. Using the representation in figure~\ref{fig:psi_partial} c), there can be one
predicate associated with the definition of a variable, and there will
be one predicate associated with the use of the variable in a $\psi$
operation. The two domains for these two predicates do not need to be
equal, only the domain of the predicate on the definition has to contain
the domain of the predicate on the $\psi$ argument. This
extension to the representation of the $\psi$ operations allows one to
perform a copy folding algorithm to remove all {\tt mov} operations in
the representation, whether they are predicated or not.

\SubSection{Psi-predicate promotion}

The extension of the $\psi$-SSA representation to the context of
partial predication brings another useful transformation to the $\psi$
operations, the $\psi$-predicate promotion.

The predicate associated with an argument in a $\psi$ operation can be
promoted, without changing the semantics of the $\psi$ operation. By
predicate promotion, we mean that a predicate can be replaced by a
predicate with a larger predicate domain. This promotion must obey
the two following conditions so that the semantics of the $\psi$
operation after the transformation is valid and unchanged.


\begin{itemize}

\item {\bf Condition 1} For an argument in a $\psi$ operation, the
domain of the predicate used on the definition of this argument must
contain the domain of the new predicate associated with this argument.

\begin{tabular}{ll}
\multicolumn{2}{l}{\it for the instructions}\\
${\tt p?}$ & ${\tt x = ...}$\\
& ${\tt y = Psi(..., q? x, ...)}$\\
\multicolumn{2}{l}{\it then}\\
& {${\tt q} \subseteq {\tt p}$}\\
\end{tabular}

\item {\bf Condition 2} For an argument in a $\psi$ operation, the
domain of the new predicate associated with it can be extended up to
include the domains of the predicates associated with arguments in the
$\psi$ operation that were defined after the definition for this
argument in the original program.

\begin{tabular}{ll}
\multicolumn{2}{l}{\it for an instruction} \\
\ \ \ \ & {$ {\tt y = Psi(p_1?x_1, p_2?x_2, ..., p_i?x_i, ..., p_n?x_n)}$} \\
\multicolumn{2}{l}{\it transformed to} \\
\ \ \ \ & {$ {\tt y = Psi(p_1?x_1, p_2?x_2, ..., p_i'?x_i, ..., p_n?x_n)}$} \\
\multicolumn{2}{l}{\it then} \\
\ \ \ \ & {${\tt p_i' \subseteq \bigcup_{k=i}^n p_k}$} \\
\end{tabular}

\end{itemize}

The first condition ensures that the $\psi$ operation is still
valid. This condition means that, in addition to predicate promotion,
speculation may have to be performed first on the definition of the
argument of the $\psi$ operation. The second condition ensures that
the value of the $\psi$ operation is not changed. We already said that
the order of the arguments in a $\psi$ operation is, from left to
right, the order, from top to bottom, of the definitions in the
control-flow dominance tree of the original program. Thus, the domain
of the predicate associated with an argument in a $\psi$ operation can
be extended up to include the domains of each of the predicates
associated with arguments at its right in the $\psi$ argument
list. With this condition, we ensure that the conditions under which
arguments at the left of the promoted argument can have there value
overridden by arguments at their right in the $\psi$ operation remain
unchanged. This condition also means that the first argument of a
$\psi$ operation can be promoted independently of the other arguments
in the $\psi$ operation, provided that the first condition is still
satisfied.

This $\psi$-predicate promotion transformation allows us to reduce the
number of predicates that need to be computed, and to reduce the dependencies
between predicate computations and conditional operations. In fact,
the first argument of a $\psi$ operation can usually be promoted under
the {\tt TRUE} predicate, provided that speculation can be
applied. Also, when disjoint conditions are computed, one of them can
be promoted to include the other conditions, usually reducing the
dependency height of the predicated expressions. The $\psi$-predicate
promotion transformation can be applied during an if-conversion
algorithm for example. A side effect of this transformation is that it
may increase the number of copy instructions to be generated during
the out of $\psi$-SSA phase, because of more live-range interference
between arguments in a $\psi$ operation, as will be explained in
the next section.

\Section{An out of Psi-SSA algorithm}

We have now described the semantics of the $\psi$ operation along with
the transformations that can be applied on it. Then, after
optimizations have been applied on a $\psi$-SSA representation, the
code must eventually be reverted back to a standard, non SSA, form. On
the SSA representation this is called the out of SSA phase. This pass
must be adapted to the $\psi$-SSA representation.

In the original paper on the $\psi$-SSA representation~\cite{Stou01},
an out of $\psi$-SSA algorithm was described. In this section,
we present a complete algorithm that extends the original algorithm to
our new representation, and also fixes one error in the original
description.

\SubSection{Conventional SSA}

The algorithm described in the original paper on $\psi$-SSA and the
algorithm we present here are both derived from the out of SSA
algorithm from Sreedhar et al.~\cite{VC+99}.

This algorithm uses $\phi$ congruence classes to create a conventional
SSA representation. Two variables {\tt x} and {\tt y} are in a
$\phi$-congruence relation if they are referenced in the same $\phi$
function, or if there exists a variable {\tt z} such that {\tt x} is
in a $\phi$-congruence relation with {\tt z} and {\tt y} is in a
$\phi$-congruence relation with {\tt z}. Then we define a $\phi$
congruence class as the transitive closure of the $\phi$-congruence
relation. The conventional SSA representation has the property that
the renaming of all the resources from a $\phi$ congruence class into
a representative name, and the elimination of the $\phi$ operations,
will not violate the semantics of the program. The Sreedhar algorithm
gives three methods, the third one being the most efficient, to
convert an SSA representation into a conventional SSA form.

\SubSection{Conventional Psi-SSA}

We define the conventional $\psi$-SSA ({\em $\psi$-CSSA}) form in a
similar way to the Sreedhar definition of the conventional SSA ({\em CSSA})
form. The congruence relation is extended to the $\psi$
operations. Two variables {\tt x} and {\tt y} are in a
$\psi$-congruence relation if they are referenced in the same $\phi$
or $\psi$ function, or if there exists a variable {\tt z} such that
{\tt x} is in a $\psi$-congruence relation with {\tt z} and {\tt y} is
in a $\psi$-congruence relation with {\tt z}. Then we define a $\psi$
congruence class as the transitive closure of the $\psi$-congruence
relation. The property of the $\psi$-CSSA form is that the renaming
into a single variable of all variables that belong to the same
congruence class, and the removal of the $\psi$ and $\phi$ operations,
results in a program with the same semantics as the original program.

Now, look at figure~\ref{fig:psi-interference} to examine the
transformations that must be performed to convert a program from a
$\psi$-SSA form into a program in $\psi$-CSSA form.

\begin{figure*}
\begin{center}
\footnotesize
\begin{tabular}{llllllll}
${\tt p?}$ & ${\tt b = ...}$ & \ \ \ \ \ \ \ \  & ${\tt p?}$ & ${\tt b = ...}$ & \ \ \ \ \ \ \ \  & ${\tt p?}$ & ${\tt b = ...}$ \\
            & ${\tt a = ...}$ & \ \ \ \ \ \ \ \  &            & ${\tt a = ...}$ & \ \ \ \ \ \ \ \  &            & ${\tt x = ...}$ \\
            &                 & \ \ \ \ \ \ \ \  & ${\tt p?}$ & ${\tt c = b}$   & \ \ \ \ \ \ \ \  & ${\tt p?}$ & ${\tt x = b}$ \\
            & ${\tt x = Psi(1?a,p?b)}$ & \ \ \ \ \ \ \ \  &   & ${\tt x = Psi(1?a,p?c)}$  & \ \ \ \ \ \ \ \ &   & \\
\\
\multicolumn{2}{l}{${\tt {\bf Psi-SSA\ form}}$} & \ \ \ \ \ \ \ \  &\multicolumn{2}{l}{${\tt {\bf Psi-CSSA\ form}}$} & \ \ \ \ \ \ \ \  &\multicolumn{2}{l}{${\tt {\bf non-SSA\ form}}$} \\
\\
            & ${\tt a = ...}$ & \ \ \ \ \ \ \ \  &            & ${\tt a = ...}$ & \ \ \ \ \ \ \ \  &            & ${\tt x = ...}$ \\
            &                 & \ \ \ \ \ \ \ \  &            & ${\tt d = a}$ & \ \ \ \ \ \ \ \  &            & ${\tt y = x}$ \\
${\tt p?}$ & ${\tt b = ...}$ & \ \ \ \ \ \ \ \  & ${\tt p?}$ & ${\tt b = ...}$ & \ \ \ \ \ \ \ \  & ${\tt p?}$ & ${\tt x = ...}$ \\
${\tt q?}$ & ${\tt c = ...}$ & \ \ \ \ \ \ \ \  & ${\tt q?}$ & ${\tt c = ...}$ & \ \ \ \ \ \ \ \  & ${\tt q?}$ & ${\tt y = ...}$ \\
           & ${\tt x = Psi(1?a,p?b)}$ & \ \ \ \ \ \ \ \  &   & ${\tt x = Psi(1?a,p?b)}$  & \ \ \ \ \ \ \ \ &   & \\
           & ${\tt y = Psi(1?a,q?c)}$ & \ \ \ \ \ \ \ \  &   & ${\tt y = Psi(1?d,q?c)}$  & \ \ \ \ \ \ \ \ &   & \\
\\
\multicolumn{2}{l}{${\tt {\bf Psi-SSA\ form}}$} & \ \ \ \ \ \ \ \  &\multicolumn{2}{l}{${\tt {\bf Psi-CSSA\ form}}$} & \ \ \ \ \ \ \ \  &\multicolumn{2}{l}{${\tt {\bf non-SSA\ form}}$} \\
\end{tabular}
\caption{$\psi$-SSA and $\psi$-CSSA forms}
\label{fig:psi-interference}
\end{center}
\end{figure*}


Looking at the first example, the dominance order of the definitions
for the variables {\tt a} and {\tt b} differs from their order from
left to right in the $\psi$ operation. Such code may appear after a
code motion algorithm has moved the definitions for {\tt a} and {\tt
b} relatively to each other. We have said that the semantics of a
$\psi$ operation is dependent on the order of its arguments, and that
the order of the arguments in a $\psi$ operation is the order of their
definitions in the dominance tree in the original program. In this
example the renaming of the variables {\tt a}, {\tt b} and {\tt x}
into a single variable will not preserve the semantics of the original
program. The order in which the definitions of the variables {\tt a},
{\tt b} and {\tt x} occur must be corrected. This is done through the
introduction of the variable {\tt c} that is defined as a copy of the
variable {\tt b}, and is inserted after the definition of {\tt
a}. Now, the renaming of the variables {\tt a}, {\tt c} and {\tt x}
into a single variable will result in the correct semantics.

In the second example, the renaming of the variables {\tt a}, {\tt b},
{\tt c}, {\tt x} and {\tt y} into a single variable will not give the
correct semantics. In fact, the value of {\tt a} used in the second
$\psi$ operation would be overridden by the definition of {\tt b}
before the definition of the variable {\tt c}. Such code will occur
after copy folding has been applied on a $\psi$-SSA representation. We
see that the value of {\tt a} has to be preserved before the
definition of {\tt b}, resulting in the code given for the $\psi$-CSSA
representation. Now, the variables {\tt a}, {\tt b} and {\tt x} can be
renamed into a single variable, and the variables {\tt d}, {\tt c} and
{\tt y} will be renamed in another variable, resulting in a program in
a non-SSA form with the correct semantics.

We will now present an algorithm that will transform a program from a
$\psi$-SSA form into its $\psi$-CSSA form. This algorithm is made of
three parts.

\begin{itemize}
\item {\bf $\psi$-normalize} This part will put all $\psi$ operations
in what we call a {\em normalized} form.
\item {\bf $\psi$-congruence} This part will grow $\psi$-congruence
classes from $\psi$ operations, and will introduce repair code where
needed.
\item {\bf $\phi$-congruence} This part will extend the
$\psi$-congruence classes with $\phi$ operations. This part is very
similar to the Sreedhar algorithm.
\end{itemize}

We detail now the implementation of each of these three parts.

\SubSection{Psi-normalize}

We define the notion of {\em normalized}-$\psi$. When $\psi$
operations are created during the construction of the $\psi$-SSA
representation, as described in~\cite{Stou01}, they are naturally
built in their normalized form. The normalized form of a $\psi$
operation has two characteristics:

\begin{itemize}
\item The predicate associated with each argument in a
normalized-$\psi$ operation is equal to the predicate used on the
unique definition of this argument.
\item The order of the arguments in a normalized-$\psi$ operation is,
from left to right, equal to the order of their definitions, from top
to bottom, in the control-flow dominance tree.
\end{itemize}

When transformations are applied to the $\psi$-SSA representation,
predicated definitions may be moved relatively to each others.
Operation speculation and copy folding may enlarge the domain of the
predicate used on the definition of a variable. These transformations
may cause some $\psi$ operations to be in a non-normalized form.

In the original algorithm described in~\cite{Stou01}, {\it
Condition 1} for the definition of the $\psi$-SSA Consistency was
identical to the second characteristic of the normalized form we
describe here. However, the original algorithm did not include a
specific normalization phase for the out of $\psi$-SSA algorithm.
There are two reasons why this step is now needed. The first reason is
that in the original $\psi$ representation, there was no predicate
associated with an argument in a $\psi$ operation. Implicitly, this
predicate was equal to the predicate used on the definition of the
argument, but these predicates can now be different in our
representation. The second reason is to fix a problem in the original
algorithm. In figure~\ref{fig:psi-copyprop}, we show an example where
copy folding on predicated code has transformed the second $\psi$
operation into a non-normalized form. The original algorithm assumed
that for variables used in the argument list of $\psi$ operations, it
was always possible to define a strict order relation between
variables in a congruence class, noted $\succ_c$. This order was
determined using the relative order of the variables in the different
$\psi$ argument lists where these variables were used. Clearly, in
this example, there is no such relation between variables {\tt a} and
{\tt b} in the non-normalized form of the $\psi$ operations.

\begin{figure}
\begin{center}
\footnotesize
\begin{tabular}{llll}
${\tt p?}$ & ${\tt a = op1}$ & ${\tt p?}$ & ${\tt a = op1}$\\
${\tt q?}$ & ${\tt b = op2}$ & ${\tt q?}$ & ${\tt b = op2}$\\
${\tt p?}$ & ${\tt c = a}$   &            &                \\
& ${\tt x = Psi(p?a,q?b)}$   & & ${\tt x = Psi(p?a,q?b)}$  \\
& ${\tt y = Psi(q?b,p?c)}$   & & ${\tt y = Psi(q?b,p?a)}$  \\
\\
\multicolumn{2}{l}{${\tt{\bf a)\ Normalized\ form}}$} & \multicolumn{2}{l}{${\tt{\bf b)\ non\ normalized\ form}}$}\\
\end{tabular}
\caption{Copy folding on $\psi$-SSA representation}
\label{fig:psi-copyprop}
\end{center}
\end{figure}

\paragraph{PSI-normalize implementation.}
A dominator tree must be available for the control-flow graph to
lookup the dominance relation between basic blocks. The dominance
relation between two operations in a same basic block will be given by
their relative positions in the basic block.

Each $\psi$ operation is processed independently. An analysis of the
$\psi$ operations in a top down traversal of the dominator tree
reduces the amount of repair code that is inserted during this pass. We
only detail the algorithm for such a traversal.

For a $\psi$ operation, the argument list is processed from left to
right. For each argument $arg_i$, the predicate associated with this
argument in the $\psi$ operation and the predicate used on the
definition of this argument are compared. If they are not equal, a new
variable is introduced and is initialized just below the definition
for $arg_i$ with a copy of $arg_i$. This definition is predicated with
the predicate associated with $arg_i$ in the $\psi$ operation. Then,
$arg_i$ is replaced by this new variable in the $\psi$
operation.

Then, we consider the dominance order of the definition for
$arg_i$, with the definition of the next argument in the $\psi$
argument list, $arg_{i+1}$. When $arg_{i+1}$ is defined on a $\psi$
operation, we recursively look for the definition of the first
argument of this $\psi$ operation, until a non-$\psi$ operation is
found. Now, if the definition we found for $arg_{i+1}$ dominates the
definition for $arg_i$, repair code is needed. A new variable is
created for this repair. This variable is initialized with a copy of
$arg_{i+1}$, guarded by the predicate associated with this argument in
the $\psi$ operation. This copy operation is inserted at the lowest
point, either after the definition of $arg_i$ or $arg_{i+1}
$\footnote{When $arg_{i+1}$ is defined by a $\psi$ operation, its
definition may appear after the definition for $arg_i$, although the
non-$\psi$ definition for $arg_{i+1}$ appears before the definition
for $arg_i$.}. Then, $arg_{i+1}$ is replaced in the $\psi$ operation
by this new variable.

The algorithm continues with the argument $arg_{i+1}$, until all
arguments of the $\psi$ operation are processed. When all arguments
are processed, the $\psi$ is in its normalized form. When all $\psi$
operations are processed, the function will contain only
normalized-$\psi$ operations.

The top-down traversal of the dominator tree will ensure that when a
variable in a $\psi$ operation is defined by another $\psi$ operation,
this $\psi$ operation has already been analyzed and put in its
normalized form. Thus the definition of its first variable already
dominates the definitions for the other arguments of the $\psi$
operation.

\paragraph{}
In figure~\ref{fig:psi-normalize} we show how this algorithm
works. The first $\psi$ operation is analyzed. The analysis starts
with argument {\tt a}. The predicate associated with this argument is
equal to the predicate used on the definition for {\tt a}, and the
definition of {\tt a} dominates the definition of {\tt b}, thus no
repair code is needed. The analysis continues with argument {\tt
b}. The predicate associated with the argument {\tt b} in the $\psi$
operation is not equal to the predicate used on the definition of {\tt
b}. A new variable {\tt e} is introduced, and is defined as a
predicated copy of {\tt b} using the predicate associated with {\tt b} in
the $\psi$ operation. Then {\tt b} is replaced by {\tt e} in this
$\psi$ operation. On the next $\psi$ operation, the definition for
{\tt c} does not dominate the definition for {\tt d}. A new variable
{\tt f} is introduced and initialized with {\tt d} under predicate
{\tt s}. This copy operation is inserted just after the definition for
{\tt c}. On the last $\psi$ operation, since {\tt y} is defined on a
$\psi$ operation we use the definition of {\tt c} as the definition
point for {\tt y}. The definition of {\tt x} does not dominate the
definition for {\tt c}, so a repair is needed. The copy {\tt g =
y} is inserted after the definition of {\tt y}, and is predicated with
the predicate associated with {\tt y} in the $\psi$ operation.

This algorithm ensures that the program contains only
normalized $\psi$ operations. This property is used by the next
two passes of the algorithm.

\begin{figure}
\begin{center}
\footnotesize
\begin{tabular}{llll}
           & ${\tt d = ...}$ &            & ${\tt d = ...}$ \\
${\tt p?}$ & ${\tt a = ...}$ & ${\tt p?}$ & ${\tt a = ...}$ \\
${\tt r?}$ & ${\tt c = ...}$ & ${\tt r?}$ & ${\tt c = ...}$ \\
           &                 & ${\tt s?}$ & ${\tt f = d}$ \\
           & ${\tt b = ...}$ &            & ${\tt b = ...}$ \\
           &                 & ${\tt q?}$ & ${\tt e = b}$ \\
${\tt p|q?}$ & ${\tt x = Psi(p?a,q?b)}$ & ${\tt p|q?}$ & ${\tt x = Psi(p?a,q?e)}$ \\
${\tt r|s?}$ & ${\tt y = Psi(r?c,s?d)}$ & ${\tt r|s?}$ & ${\tt y = Psi(r?c,s?f)}$ \\
             &                          & ${\tt r|s?}$ & ${\tt g = y}$ \\
           & ${\tt z = Psi(p|q?x,r|s?y)}$ &            & ${\tt z = Psi(p|q?x,r|s?g)}$ \\
\end{tabular}
\caption{Converting $\psi$ operations into their normalized form }
\label{fig:psi-normalize}
\end{center}
\end{figure}

\SubSection{Psi-congruence}

In this pass, we repair the $\psi$ operations when variables
cannot be put into the same congruence class, because their live ranges
interfere. In the same way as Sreedhar gave a definition of the
liveness on the $\phi$ operation, we first give a definition for the
liveness on $\psi$ operations. With this definition of liveness, an
interference graph is built.

\paragraph{Liveness and interferences in Psi-SSA.}
We have already seen that in some cases, repair code is needed so that the
arguments and definition of a $\psi$ operation can be renamed into a
single name. Here, we give a definition of the liveness on $\psi$
operations such that these cases can be easily detected by observing that
live-ranges for variables in a $\psi$ operation overlap. Our
definition of liveness differs from the definition used in the
original paper, and allows for more precise detection and repair of
the interferences between variables in $\psi$ operations.

Consider the code in figure~\ref{fig:psi-select}. The $\psi$ operation
has been replaced by explicit {\tt select} operations on each
predicated definition. In this example, there is no relation between
predicates {\tt p} and {\tt q}. Each of these {\tt select} operations
makes an explicit use of the variable immediately to its left in the
argument list of the original $\psi$ operation. We can see that a
renaming of the variables {\tt a}, {\tt b}, {\tt c} and {\tt x} into a
single representative name will still compute the same value for the
variable {\tt x}. Note that this transformation can only be performed
on normalized $\psi$ operations, since the definition of an argument
must be dominated by the definition of the argument immediately to its
left in the argument list of the $\psi$ operation. Using this
equivalent representation for the $\psi$ operation, we now give a
definition of the liveness for the $\psi$ operations.

{\bf Definition} {\em We say that the point of use of an argument in a
normalized $\psi$ operation occurs at the point of definition of the
argument immediately to its right in the argument list of the $\psi$
operation. For the last argument of the $\psi$ operation, the point of
use occurs at the $\psi$ operation itself.  }

\begin{figure}
\begin{center}
\footnotesize
\begin{tabular}{llll}
           & ${\tt a = op1}$ & ${\tt a = op1}$ \\
${\tt p?}$ & ${\tt b = op2}$ & ${\tt b = p\ ?\ op2\ :\ a}$ \\
${\tt q?}$ & ${\tt c = op3}$ & ${\tt c = q\ ?\ op3\ :\ b}$ \\
           & ${\tt x = Psi(1?a,p?b,q?c)}$ & ${\tt x = c}$ \\
\\
\multicolumn{2}{l}{\tt {\bf a) Psi-SSA form}} & {\tt {\bf b) select form}}
\end{tabular}
\caption{$\psi$ and select operations equivalence}
\label{fig:psi-select}
\end{center}
\end{figure}

Given this definition of liveness on $\psi$ operations, and using the
definition of liveness for $\phi$ operations given by Sreedhar, a
traditional liveness analysis can be run. Then an interference graph
can be built to collect the interferences between variables involved
in $\psi$ or $\phi$ operations.

\paragraph{Repairing interferences on $\psi$ operations.}
We now present an algorithm that creates congruence classes with
$\psi$ operations such that there are no interference between two
variables in the same congruence class.

First, the congruence classes are initialized such that each variable
in the $\psi$-SSA representation belongs to its own congruence
class. Then, $\psi$ operations are processed one at a time, in
no specific order. Two arguments of a $\psi$ operation interfere if
one or more variables from the congruence class of the first argument and one or more
variables from the congruence class of the second argument
interfere. When there is an interference, the two $\psi$ arguments are
marked as needing a repair. When all pairs of arguments of the $\psi$
operation are analyzed, repair code is inserted. For each argument in
the $\psi$ operation that needs a repair, a new variable is
introduced. This new variable is initialized with a predicated copy of
the argument's variable. The copy operation is inserted just below the
definition of the argument's variable, predicated with the predicate
associated with the argument in the $\psi$ operation.

Once a $\psi$ operation has been processed, the interference graph
must be updated, so that other $\psi$ operations are correctly
handled. Interferences for the newly introduced variables must be
added to the interference graph. Conservatively, we can say that each
new variable interferes with all the variables that the
original variable interfered with, except those variables that
are now in its congruence class. Also, conservatively, we can say
that the original variable interferes with the new variable in order
to avoid a merge of a later $\psi$ or $\phi$ operation of the two
congruence classes these two variables belong to. The conservative
update of the interference graph may increase the number of copies
generated during the conversion to the $\psi$-CSSA form.

Consider the code in figure~\ref{fig:live-interference} to see how this
algorithm works. The definition of liveness on the $\psi$ operation
will create a live-range for variable {\tt a} that extends down to the
definition of {\tt b}, but not further down. Thus, the variable {\tt a}
does not interfere with the variables {\tt b}, {\tt c} or {\tt x}. The
live-range for variable {\tt b} extends down to its use in the
definition of variable {\tt d}. This live-range creates an
interference with the variables {\tt c} and {\tt x}. Thus variables
{\tt b}, {\tt c} and {\tt x} cannot be put into the same congruence
class. These variables are renamed respectively into variables {\tt
e}, {\tt f} and {\tt g} and initialized with predicated copies. These
copies are inserted respectively after the definitions for {\tt b},
{\tt c} and {\tt x}. Variables {\tt a}, {\tt e}, {\tt f} and {\tt g}
can now be put into the same congruence class, and will be renamed
later into a unique representative name.

\begin{figure}
\begin{center}
\footnotesize
\begin{tabular}{llll}
${\tt p?}$ & ${\tt a = ...}$ & ${\tt p?}$ & ${\tt a = ...}$ \\
${\tt q?}$ & ${\tt b = ...}$ & ${\tt q?}$ & ${\tt b = ...}$ \\
           &                 & ${\tt q?}$ & ${\tt e = b}$ \\
${\tt r?}$ & ${\tt c = ...}$ & ${\tt r?}$ & ${\tt c = ...}$ \\
           &                 & ${\tt r?}$ & ${\tt f = c}$ \\
           & ${\tt x = Psi(p?a,q?b,r?c)}$ & & ${\tt g = Psi(p?a,q?e,r?f)}$ \\
           &                 &            & ${\tt x = g}$ \\
${\tt s?}$ & ${\tt d = b+1}$ & ${\tt s?}$ & ${\tt d = b+1}$ \\
\end{tabular}
\caption{Elimination of $\psi$ live-interference}
\label{fig:live-interference}
\end{center}
\end{figure}

\SubSection{Phi-congruence}

When all $\psi$ operations are processed, the congruence classes built
from $\psi$ operations are extended to include the variables in $\phi$
operations. In this part, the algorithm from Sreedhar is used, with a few
modifications.

The first modification is that the congruence classes must not be
initialized at the beginning of this process. They have already been
initialized at the beginning of the $\psi$-congruence step, and were
extended during the processing of $\psi$ operations. These congruence
classes will be extended now with $\phi$ operations during this step.

The other modification is that the live-analysis run for this part
must also take into account the special liveness rule on the $\psi$
operations. The reason for this is that for any two variables in the
same congruence class, any interference, either on a $\psi$ or on a
$\phi$ operation, will not preserve the correct semantics if the
variables are renamed into a representative name.

All other parts of the algorithm are unchanged, and in particular, any
of the three algorithms described for the conversion into a CSSA form
can be used.

We have described a complete algorithm to convert a $\psi$-SSA
representation into a $\psi$-CSSA representation. The final step to
convert the code into a non-SSA form is a simple renaming of all the
variables in the same congruence class into a representative name. The
$\psi$ and $\phi$ operations are then removed.

Now that a complete algorithm has been described to convert a $\psi$-SSA
representation to a $\psi$-CSSA representation, we will present some
improvements that can be added so as to reduce the number of copies
inserted by this algorithm.

\SubSection{Improvements to the out of Psi-SSA algorithm}

Below we present a list of improvements that can be added to the
algorithm.

\paragraph{Non-normalized $\psi$ operations with disjoint predicates.}
When two arguments in a $\psi$ operation do not have their definitions
correctly ordered, the $\psi$ operation is not normalized. We
presented an algorithm to restore the normalized property by adding a
new predicated definition of a new variable. However, if we know that
the predicate domains of the two arguments are actually disjoint, the
semantics of the $\psi$ operation is independent on their relative
order. So, instead of adding repair code, these two arguments can
simply be reordered in the $\psi$ operation itself, to restore the
normalized property.

\paragraph{Interference with disjoint predicates.}
When the live-ranges of two variables overlap, an interference is
added for these two variables in the interference graph. If the
definitions for these variables are predicated definitions, their
live-ranges are only valid under a specific predicate domain. These
domains are the domains of the predicates used on the definitions of
the variables. Then, if these domains are disjoint, then although the
live-range overlap, they are on disjoint conditions and thus they do
not create an interference in the interference graph. Removing this
interference from the interference graph will avoid the need to add repair code
when live-ranges on disjoint predicates overlap.

\paragraph{Repair interference on the left argument only.}

When an interference is detected between two arguments in a $\psi$
operation, only the argument on the left actually needs a repair. The
reason is that, since the $\psi$ operations are normalized, the
definition of an argument is always dominated by the definition of an
argument on its left. Thus adding a copy for the argument on the right
will not remove the interference.

\paragraph{Interference with the result of a $\psi$ operation.}

When the live-range for an argument of a $\psi$ operation overlaps
with the live-range of the variable defined by the $\psi$ operation,
this interference can be ignored. Actually, there are two cases to
consider:

\begin{itemize}
\item If the argument is not the last one in the $\psi$ operation, and
its live-range overlaps with the live-range of the definition of the
$\psi$ operation, then this live-range also overlaps with the
live-range of the last argument. Thus this interference will already
be detected and repaired.

\item If the argument is the last one of the $\psi$ operation, then
the value of the $\psi$ operation is the value of this last argument,
and this argument and the definition will be renamed into the same
variable out of the SSA representation. Thus, there is no need to
introduce a copy here.
\end{itemize}

\Section{Experimental results}

The $\psi$-SSA representation has been implemented in our production
compiler for the ST200 family processors~\cite{faraboschi00lx}. This
compiler is based on the Open64 compiler technology, and the
$\psi$-SSA representation has been used to implement optimizations in
the code generator part of the compiler. The experiment has been conducted
on a variant of the ST231 processor. The ST231 is a 4-issue VLIW
processor that targets multimedia and digital consumer embedded
applications. It is composed of four 32-bit integer ALUs, two 32x32
multipliers, one load/store unit, a branch unit, 64 32-bit general
purpose registers and 8 1-bit branch registers. The variant we used includes
support for partial predication, through predicated load and store
instructions and a {\tt select} instruction.

The $\psi$-SSA representation is used in the backend of our compiler
to implement several optimizations. These optimizations include a
range-propagation analysis to remove redundant or useless
computations, an address expressions analysis to optimize the use of
available addressing modes, and an if-conversion
algorithm~\cite{Bru06}. However, these transformations will only very
occasionally produce non-normalized $\psi$ operations or add
interferences between variables in $\psi$ operations.

In order to analyze the situations where some repair code is
introduced on $\psi$ operations, we added a copy folding algorithm
just before the out of $\psi$-SSA algorithm. We ran our algorithms on
a set of small benchmarks from multimedia applications. The results
are reported in figures~\ref{fig:nopromot} and~\ref{fig:promot}. In
these experiments we measured the number of copy operations that were
inserted during each of the three steps of the out of $\psi$-SSA
algorithm, and we measured the total number of copy operations in the
program after the out of $\psi$-SSA phase.

In figure~\ref{fig:nopromot}, we report the figures when no
$\psi$-predicate promotion algorithm was applied. In the first column,
we report the number of copies when the if-conversion and the copy
folding optimizations are not run. As expected, the transformations
that are performed on the SSA representations do not break the
$\psi$-SSA conventional property on these benchmarks, which results in
no copy operation being inserted during the out of $\psi$-SSA
phase. The second column shows the results when the if-conversion
transformation is performed. A number of copy operations are inserted
during the $\psi$-normalize step, which shows that the if-conversion
algorithm generates non-normalized $\psi$ operations. Most of these
non-normalized $\psi$ operations are due to the predicate being
different on the definition of the variable and on its use in the
$\psi$ operation. The $\psi$-congruence step creates no additional
copy operations, which means that no interference was detected between
variables on $\psi$ operations during this step. In the third column,
copy folding was performed in addition to the if-conversion
transformation. This resulted in additional non-normalized $\psi$
operations. These additional non-normalized operations are created
when predicated copy operations are folded, resulting in more $\psi$
operations with a different predicate on the definition for a variable
and its use in the $\psi$ operation. There is also one interference in
the $\psi$-congruence step that was created by the copy folding. This
copy operation cannot be optimized away. The large number of copy
operations generated during the $\phi$-congruence step is mostly due
to the fact that we only implemented the second method of the Sreedhar
algorithm, use of the third method would reduce this
number. Finally, the total number of copy instructions after the out
of $\psi$-SSA phase is greater after copy folding has been performed,
mostly due to the number of copy operations generated during the
$\phi$-congruence step.

In figure~\ref{fig:promot}, we report the figures when
$\psi$-predicate promotion algorithm was performed. The
$\psi$-predicate promotion propagates into the $\psi$ operations the
effect of the speculation that was performed during the if-conversion
algorithm. The main reason to perform the $\psi$-predicate promotion is to
reduce the number of predicates that must be computed in the code. This
transformation also reduces the number of non-normalized $\psi$
operations, so that fewer copy operations need to be inserted during the
$\psi$-normalize step. This is shown in the second and third columns
for the $\psi$-normalize step. The number of copy instructions
introduced in this step is reduced compared to the number of copy
instructions that were introduced in the same step without the
$\psi$-predicate promotion. On the $\psi$-congruence step, we see that
performing the $\psi$-predicate promotion actually increased the number
of interferences to be repaired. In fact, these interferences also
existed without the $\psi$-predicate promotion, but, due to the
smaller number of non-normalized $\psi$-operations, they were not
repaired as a side effect of the $\psi$-normalize step.

On the last line of this figure, we see that after the out of
$\psi$-SSA phase there is a small decrease in the number of copy
instructions in the code when $\psi$-predicate promotion is
performed. The cases where fewer copy operations are generated occur
in loops where a $\psi$ operation uses and defines variables that are
used in the same $\phi$ operation. Such a situation is described in
figure~\ref{fig:interferences}. The $\psi$ operation in the code on
the left is not normalized, because the predicate for the variable
{\tt c} is different on its definition in the $\phi$ operation and on
its use in the $\psi$ operation. In the code on the right, a variable
{\tt e} has been added to normalize this $\psi$ operation. The
$\psi$-congruence step creates a congruence class with the variables
{\tt e}, {\tt d} and {\tt b}, since there is no interference between
these variables. The $\phi$ operation is then processed during the
$\phi$-congruence step. The interferences between the variable {\tt c}
and the variables in the congruence class for {\tt b} are checked. In
fact, the variables {\tt c} and {\tt e} interfere, which will require
that a new variable is introduced and a new copy instruction is
inserted. When the predicate promotion is performed first on the
$\psi$ operation for the variable {\tt c}, the variable {\tt e} is no
longer introduced. The $\psi$-congruence step creates a congruence
class with variables {\tt c}, {\tt d} and {\tt b}. In the
$\phi$-congruence step, when processing the $\phi$ operation, no
interference needs to be repaired since the variables {\tt b} and {\tt
c} are already in the same congruence class, and thus no additional
copy instruction is inserted.

Future work will include improving the out of SSA algorithm in order
to reduce the number of copies generated during this phase. In
particular, we will work on a better integration between the
$\psi$-congruence and the $\phi$-congruence steps to avoid the cases
where repair code introduced in the $\psi$-congruence step creates
interferences to be repaired in the $\phi$-congruence step.

\begin{figure}
\begin{center}
\footnotesize
\begin{tabular}{|l|r|r|r|}
\hline
                & no \ \  &         & if-conv \\
\ \ \ \  copies & if-conv & if-conv & + folding \\

\hline
psi-normalize & +0 & +129 & +163 \\
psi-congruence & +0 & +0 & +1 \\
phi-congruence & +0 & +82 & +1543 \\
\hline
total copies & 7041 & 7268 & 7948 \\
\hline
\end{tabular}
\caption{Out of $\psi$-SSA without $\psi$-predicate promotion}
\label{fig:nopromot}
\bigskip
\bigskip
\begin{tabular}{|l|r|r|r|}
\hline
                & no \ \  &         & if-conv \\
\ \ \ \  copies & if-conv & if-conv & + folding \\

\hline
psi-normalize & +0 & +34 & +65 \\
psi-congruence & +0 & +5 & +11 \\
phi-congruence & +0 & +11 & +1493 \\
\hline
total copies & 7041 & 7107 & 7810 \\
\hline
\end{tabular}
\caption{Out of $\psi$-SSA with $\psi$-predicate promotion}
\label{fig:promot}
\end{center}
\end{figure}

\begin{figure}[ht]
\footnotesize
\begin{tabular}{lllll}
         & ${\tt loop:}$          & \  &          & ${\tt loop:}$ \\
         & ${\tt c = Phi(a, b)}$  &          &          & ${\tt c = Phi(a, b)}$ \\
         &                      &          & ${\tt \overline{p} ?}$ & ${\tt e = c}$ \\
         & ${\tt q = c < 10}$     &          &          & ${\tt q = c < 10}$ \\
${\tt p?}$ & ${\tt d = op1}$        &          & ${\tt p?}$ & ${\tt d = op1}$ \\
         & ${\tt b = Psi(\overline{p}?c, p?d)}$ & & & ${\tt b = Psi(\overline{p}?e, p?d)}$ \\
${\tt q?}$ & ${\tt goto\ loop}$      &          & ${\tt q?}$ & ${\tt goto\ loop}$ \\
\end{tabular}
\caption{$\psi$-normalize adds new interferences}
\label{fig:interferences}
\end{figure}

\Section{Conclusion}

In this article we presented several aspects of the $\psi$-SSA
representation. The $\psi$-SSA representation is an extension of the
SSA representation to support predicated code, where some definitions
are conditionally executed depending on the value of a guard
register. We presented an improvement to the original $\psi$-SSA
representation to better support architectures with only a partially
predicated instruction set. We added a new transformation that can be
performed on the $\psi$ operations in the context of partial
predication, namely the $\psi$-predicate promotion, which is useful
for example in an if-conversion algorithm. Finally, we presented a
detailed implementation of the out of $\psi$-SSA representation
algorithm, which includes the support for partially predicated
architectures and fixes an error in the original algorithm. The
$\psi$-SSA representation is implemented in our production compiler
for the ST200 family processors, and is used to perform several
algorithms on the $\psi$-SSA representation, including an
if-conversion optimization.

\Section{Acknowledgements}

I would like to thank Christian Bruel for his implementation of an
if-conversion algorithm under $\psi$-SSA, Christophe Guillon and
Fabrice Rastello for their very useful remarks and feedback, and
Stephen Clarke for his review.


%

\bibliographystyle{latex8}
\bibliography{psi-ssa}

\begin{thebibliography}{10}\setlength{\itemsep}{-1ex}\small

\bibitem{Bru06}
C.~Bruel.
\newblock If-conversion ssa framework for partially predicated vliw
  architectures.
\newblock SIGPLAN, ACM and IEEE, March 2006.

\bibitem{CCK+97}
F.~Chow, S.~Chan, R.~Kennedy, S.-M. Liu, R.~Lo, and P.~Tu.
\newblock A new algorithm for partial redundancy elimination based on ssa form.
\newblock {\em ACM SIG{\-}PLAN Notices}, 32(5):273 -- 286, 1997.

\bibitem{ICCC::ChowCLLS1996}
F.~Chow, S.~Chan, S.-M. Liu, R.~Lo, and M.~Streich.
\newblock Effective representation of aliases and indirect memory operations in
  {SSA} form.
\newblock In T.~Gyimothy, editor, {\em Compiler Construction, 6th International
  Conference}, volume 1060 of {\em Lecture Notes in Computer Science}, pages
  253--267, Link{\"o}ping, Sweden, 24--26~Apr. 1996. Springer.

\bibitem{Cli95}
C.~Click.
\newblock Global code motion global value numbering.
\newblock In {\em SIGPLAN International Conference on Programming Languages
  Design and Implementation}, pages 246 -- 257, 1995.

\bibitem{CFR+91}
R.~Cytron, J.~Ferrante, B.~Rosen, M.~Wegman, and K.~Zadeck.
\newblock Efficiently computing static single assignment form and the control
  dependence graph.
\newblock {\em ACM Transactions on Programming Languages and Systems},
  13(4):451 -- 490, 1991.

\bibitem{Fang96}
J.~Fang.
\newblock Compiler algorithms on if-conversion, speculative predicates
  assignment and predicated code optimizations.
\newblock In {\em 9th International Workshop on Languages and Compilers for
  Parallel Computing (LCPC), LNCS~\#1239}, pages 135 -- 153, 1996.

\bibitem{faraboschi00lx}
P.~Faraboschi, G.~Brown, J.~A. Fisher, G.~Desoli, and F.~Homewood.
\newblock {Lx}: a technology platform for customizable {VLIW} embedded
  processing.
\newblock In {\em The 27th Annual International Symposium on Computer
  architecture 2000}, pages 203--213, New York, NY, USA, 2000. ACM Press.

\bibitem{KS98}
K.~Knobe and V.~Sarkar.
\newblock Array ssa form and its use in parallelization.
\newblock In {\em ACM Symposium on the Principles of Programming Languages},
  pages 107 -- 120, 1998.

\bibitem{morgan98}
R.~Morgan.
\newblock Building an optimizing compiler, January 1998.

\bibitem{VC+99}
V.~Sreedhar, R.~Ju, D.~Gillies, and V.~Santhanam.
\newblock Translating out of static single assignment form.
\newblock In {\em Static Analysis Symposium, Italy}, pages 194 -- 204, 1999.

\bibitem{Stou01}
A.~Stoutchinin and F.~{de Ferri\`ere}.
\newblock Efficient static single assignment form for predication.
\newblock In {\em 34th annual ACM/IEEE international symposium on
  Microarchitecture}, pages 172--181. IEEE Computer Society, 2001.

\bibitem{Stou04}
A.~Stoutchinin and G.~Gao.
\newblock If-conversion in {SSA} form.
\newblock In {\em Proceedings of Euro-Par 2004 Parallel Processing}, volume
  3149 of {\em LNCS}, pages 336--345. SIGPLAN, Springer Verlag, Aug 2004.

\bibitem{WZ91}
M.~Wegman and K.~Zadeck.
\newblock Constant propagation with conditional branches.
\newblock {\em ACM Transactions on Programming Languages and Systems},
  13(2):181 -- 210, 1991.

\bibitem{Wolfe92}
M.~Wolfe.
\newblock Beyond induction variables.
\newblock In {\em SIGPLAN International Conference on Programming Languages
  Design and Implementation}, pages 162 -- 174, 1992.

\end{thebibliography}

\end{document}